\documentclass[12pt]{article}

\begin{document}
\pagestyle{empty}
\begin{center}
 {\Large \bf  The  analytical singlet 
$\alpha_s^4$ QCD contributions into the  $e^+e^-$-annihilation 
Adler function and   the generalized Crewther relations}\\ 
\vspace{0.1cm}
{\bf A.L. Kataev\footnote{E-mail:kataev@ms2.inr.ac.ru} }\\
\vspace{0.1cm}
Institute for Nuclear Research of the Academy of Sciences of
Russia,\\60th October Anniversary Prospect 7a,\\ 117312, Moscow, Russia\\
\end{center}
\begin{center}
{\bf ABSTRACT}
\end{center}
\noindent
The generalized Crewther relations in the channels of the 
non-singlet and vector  quark currents are considered. 
These relations follow from the double application of the   operator product 
expansion approach to the same 
axial vector-vector-vector triangle  amplitude in two regions,    
adjoining to  the angle   sides  
$(x,y)$ (or $p^2,q^2$). 
We assume  that the generalized Crewther relations in 
these two kinematic regimes result in the 
existence of the same perturbation expression for two 
products of the coefficient functions  
of annihilation and deep-inelastic scattering 
processes in the non-singlet and vector channels.   
This feature  explains   
the conformal symmetry motivated  cancellations   
between   the singlet  $\alpha_s^3$ corrections 
to  the Gross-Llewellyn Smith sum  rule $S_{GLS}$ of ${\nu N}$ 
deep inelastic scattering and the singlet $\alpha_s^3$ correction    
to  the $e^+e^-$ annihilation 
Adler function $D_A^{V}$ in the product 
of the corresponding  perturbative series.  Taking into account  
the  Baikov-Chetyrkin-Kuhn 
4-th order  result  for $S_{GLS}$  
and the perturbative  effects of 
the violation of the conformal symmetry in the generalized Crewther relation,  
we obtain the   analytical 
contribution to the singlet  $\alpha_s^4$ correction to the  
$D_A^{V}$-function.   
Its  a-posterior comparison with the recent result of 
direct diagram-by-diagram evaluation of the singlet
4-th order corrections 
to $D_A^{V}$- function demonstrates the coincidence  of the predicted  
and obtained  $\zeta_3^2$-contributions to  the singlet term.  They 
can be  obtained 
in the conformal invariant limit   from the original Crewther relation. 
Therefore, on the contrary to previous belief, the  appearance 
of $\zeta_3$-terms in the perurbative series in 
quantum field theory gauge models   
does not contradict to the property of the  the conformal symmetry 
and  can be considered as  regular feature.
The Banks-Zaks motivated 
relation between   our  predicted and the  obtained directly  
4-th order 
corrections  is mentioned.  It   
confirms  the expectation, previously made by Baikov-Chetykin-Kuhn,  
that at the 5-loop level the generalized Crewther relation in the channel 
of vector currents may receive additional singlet contribution, which in 
this order of perturbation theory is  proportional 
to the first coefficient of the QCD $\beta$-function.

\noindent
\\[1cm]

This work is dedicated to K. G.  Chetyrkin on the  occasion of his 
60th anniversary.\\ [1cm]


\setcounter{page}{1}
\pagestyle{plain}

Quite recently the 4-th order perturbative coefficients to the 
flavour non-singlet (NS) part of the $e^+e^-$-annihilation Adler 
D-function $D_A^{NS}$
and to the NS Bjorken sum rule $S_{Bjp}$ of the deep-inelastic 
scattering (DIS) process  of polarised 
leptons on nucleons  were evaluated symbolically within general 
$SU(N_c)$ colour gauge 
group \cite{Baikov:2010je}. The explicit analytical 
expression for the 4-th  perturbative  
coefficient of $S_{Bjp}$ in the $\overline{MS}$-scheme   was obtained     
in Ref. \cite{Kataev:2010du} 
by inverting the order $a_s^4$ analytical results for $1/S_{Bjp}$, 
presented in Ref.\cite{Baikov:2010je}.
Apart of the phenomenological   
applications to  the  analysis   of the  experimental data  for the  
semi-hadron 
width of $\tau$-lepton  \cite{Davier:2008sk}\footnote{Note, that in 
this analysis the $SU(3)$  variant
\cite{Baikov:2008jh} 
of the $SU(N_c)$   expression   for $D_A^{NS}$ \cite{Baikov:2010je} was used.}
and of  the experimental data for the Bjorken polarised sum rule \cite{Khandramai:2011zd},
the $\overline{MS}$-scheme results of Ref.\cite{Baikov:2010je} 
have also  important   theoretical 
consequences.  As  was emphasised in Refs.\cite{Baikov:2010je},
\cite{Kataev:2010du},
the analytical  $SU(N_c)$-group expressions of Ref.\cite{Baikov:2010je} 
play essential  
role in the  verification of the discovered in Ref. \cite{Broadhurst:1993ru}
$\beta$-function 
factorisable  
representations for  the conformal symmetry breaking  term in the  
QCD generalization of the  original 
quark-parton model  Crewther   relation  \cite{Crewther:1972kn}.  
The validity of this    discovery of              
Ref. \cite{Broadhurst:1993ru} was proved in the $\overline{MS}$-scheme in all 
orders of perturbation theory \cite{Crewther:1997ux} (for some additional discussions 
see the   review  of Ref. 
\cite{Braun:2003rp}). 

It is worth to stress,   that  to apply self-consistently  the 
4-th order perturbative  results of Ref. \cite{Baikov:2010je} in  
the  analysis of the $e^+e^-\rightarrow{hadrons}$ data  above thresholds of 
charm quarks production     and in the fits of  the precise LEP data  
for  $Z^0\rightarrow 
{hadrons}$ decay width as well,  it is necessary to find  still unknown  
4-th order singlet (SI) contribution to the Adler functions $D_A$ of 
vector quark currents. In the previous $a_s^3$ 
order of perturbation theory  these type of   
corrections were   analytically evaluated  in 
Ref. \cite{Gorishnii:1990vf} and confirmed later on in 
\cite{Surguladze:1990tg} and \cite{Chetyrkin:1996ez}. 

In this work the  symbolical  expressions for the 4-th order SI  
contributions to $D_A$-function are predicted. 
Three important theoretical inputs  will be used:
\begin{itemize}
\item 
the universality of the original  Crewther relation \cite{Crewther:1972kn}, 
which is valid in the conformal invariant limit not only  for the   
product of the coefficient functions of 
$D_A^{NS}$-function   and NS DIS  Bjorken sum rule, 
but for the product of the coefficient functions $D_A^{V}$ 
of the  Adler D-function 
of vector currents and 
of the coefficient function of the Gross-Llewellyn DIS sum rule 
for  neutrino-nucleon DIS
as well
\cite{Adler:1973kz}; 
\item   
the statement of  Ref.\cite{Kataev:1996ce}   
that the product of the normalised perturbative  
coefficient functions 
of the    vector currents D-function  and of the  
Gross-Llewellyn sum rule  is identical to  
to   the $\overline{MS}$-scheme 
QCD expression  of the generalized    Crewther relation in   
the NS  case;  
\item 
the obtained in the case of the general $SU(N_c)$ group 
analytical expressions  for  
the $O(a_s^4)$ flavour SI
perturbative contributions  
to the Gross-Llewellyn Smith  sum rule 
in the $\overline{MS}$-scheme \cite{Baikov:2010iw} 
and the previous $O(a_s^3)$ order scheme-independent   SI   
correction  to this sum rule  which  
was  analytically  evaluated in Ref.\cite{Larin:1991tj}.
\end{itemize} 
         
Consider now physical quantities to be analysed in this work. 
Perturbative expression for  the $e^+e^-$  
Adler D-function in the vector channel has the following form 
\begin{equation}
\label{DV}
D_A^{V}(a_s)= Q^2\int_0^{\infty}\frac{R^{e^+e^-}(s)}{(s+Q^2)^2}ds= 
D_A^{NS}(a_s)+D_A^{SI}(a_s)  
\end{equation}
where the functions in the r.h.s. of Eq.(\ref{DV}) 
are defined as  
\begin{equation}
\label{DANS}
D_A^{NS}(a_s)=d_R\bigg(\sum_F Q_F^2\bigg) C_A^{NS}(a_s)~~, ~~~
D_A^{SI}(a_s)= \bigg(\sum_F Q_F \bigg)^2 C_A^{SI}(a_s)~~~.
\end{equation}         
Here  $Q_f$ are the  electric   charges of quarks  
and  $d_R$ is the dimension of $SU(N_c)$-group representation 
(for detailed 
definitions and studies  see e.g. Ref.\cite{vanRitbergen:1998pn} ). 
It enters in the normalisation factor of the NS contribution to the 
Adler vector function  and do not enter in the normalisation 
factor of the SI part  of  
Eq.(\ref{DV}).
 
Within the 
fundamental representation of $SU(N_c)$ group one has  $d_R=N_c$. 
The coefficient functions   $C_A^{NS}$ and $C_A^{SI}$ have 
the following perturbation expansions 
\begin{equation}
\label{CASI}
C_A^{NS}(a_s)=  \bigg(1+\sum_{n\geq 1}d_n^{NS}a_s^n\bigg)~~~,~~~  
C_A^{SI}(a_s)=   \sum_{n\geq 3} d_n^{SI}a_s^n    ~~~.
\end{equation}
Within perturbation theory the polarised Bjorken   DIS sum rule 
is defined as   
\begin{equation}
S_{Bjp}^{NS}(a_s)=\int_0^1 g_1^{lp-ln}(x,Q^2)dx=\frac{g_a}{6}C_{Bjp}^{NS}(a_s)
\end{equation}
where the coefficient function can be expressed  as 
\begin{equation}
C_{Bjp}^{NS}(a_s)=\bigg(1 +\sum_{m\geq 1} c_m^{NS}a_s^m\bigg )
\end{equation}
The known   
coefficients $d_1$-$d_4$ and $c_1$-$c_4$ contain the powers of  
$SU(N_c)$ colour 
group structures $C_F$, $C_A$, $T_FN_F$ and $d^{abcd}$
\cite{Baikov:2010je}, 
where $C_F$ and $C_A$ are the Casimir operators, $N_F$ is the number of flavours, 
$T_F$ is the normalisation factor of the trace of product of two $SU(N_c)$ 
generators  and $d^{abcd}$ are  the structure constants of $SU(N_C)$-group.

The Gross-Llewellyn Smith sum rule of $\nu N$ DIS can be defined as  
\begin{equation}
S_{GLS}(a_s)=\frac{1}{2}\int_0^1 F_3^{\nu p+\overline{\nu}p}(x,Q^2)dx=
3C_{GLS}(a_s)+O(\frac{1}{Q^2})
\end{equation} 
where its pure perturbative coefficient function can be decomposed 
to the NS and SI parts as 
\begin{equation}
\label{GLS}
C_{GLS}(a_s) = C_{GLS}^{NS}(a_s)+C_{GLS}^{SI}(a_s)~~. 
\end{equation}
Note, that for the NS part the following identity holds: 
\begin{equation}
\label{NSGLS}
C_{GLS}^{NS}(a_s)=C_{Bjp}^{NS}(a_s)
\end{equation}
The validity of Eq.(\ref{NSGLS}) was  first  proved within 
dimensional regularization in Ref.\cite{Gorishnii:1985xm} by demonstration 
that the    
l.h.s and r.h.s. of Eq.(\ref{NSGLS}) coincide    
after proper definition of axial Ward identities
in QCD. The  necessity of performing  extra finite renormalisation 
in the process of evaluation  of       
2-loop corrections to dimensionally regularized  
Adler $D$-functions of axial-vector  
currents was realized  in the independent, but simultaneous works 
of Refs.\cite{Antoniadis:1979us,Trueman:1979en}. 

As was demonstrated in Ref.\cite{Larin:1991tj},  
at the 3-rd order of perturbation theory the coefficient 
function of  the Gross-Llewellyn Smith sum rule of Eq.(\ref{GLS})
starts to differ from 
$C_{Bjp}^{NS}(a_s)$ due to the  appearance of  the additional SI 
contributions.  
They  enter into the expression of the  following coefficient function~:  
\begin{equation}
\label{GLSSI}
C_{GLS}^{SI}(a_s)=  \sum_{n\geq 3} c_n^{SI}a_s^n~~~. 
\end{equation}
Its  first scheme-independent  term has the 
following analytical form \cite{Larin:1991tj} 
\begin{equation}
c_3^{SI}=\frac{d^{abc}d^{abc}N_F}{d_R}c_{3,1}^{SI}=\frac{d^{abc}d^{abc}N_F}{d_R} 
\bigg(-\frac{11}{192}+\frac{1}{8}\zeta_3\bigg)~~~~.
\end{equation}
Notice, the appearance of $SU(N_c)$ group structure 
$d^{abc}d^{abc}$. 
Next  coefficient  
 was analytically 
evaluated in the $\overline{MS}$ scheme in Ref.\cite{Baikov:2010iw}. It  
has the following representation:
\begin{equation}
c_4^{SI}=~\frac{d^{abc}d^{abc}N_F}{d_R}\bigg(C_F~c_{4,1}^{SI}+C_A~c_{4,2}^{SI}
+T_FN_F~c_{4,3}^{SI}\bigg)
\end{equation}
where all coefficients were analytically  evaluated in Ref.\cite{Baikov:2010iw}. 

Consider now  the concrete   
consequences  of  the  conformal symmetry 
and of the effects of its   violation  
in the theory of strong interactions.  
The validity of the  conformal invariance    in the quark-parton model
leads to the existence of the     
Crewther relation \cite{Crewther:1972kn}. 
In the 
renormalizable gauge quantum field models (say QED or QCD)
this relation   
can be written down as   
\begin{equation}
C_A^{NS}(a_s)\times  C_{Bjp}^{NS}(a_s)|_{c-i}  = 1~~~.
\end{equation}
Here, due to the property of the 
the conformal invariance,   $a_s$  does      
not depend on transferred momentum $Q^2$.

Note, that  
the similar Crewther relation is also valid  
for the product of the $D_A^{V}$-  
and the Gross-Llewellyn Smith sum rule coefficient functions: 
\begin{equation}
\label{CDI}
C_A^{V}(a_s)\times C_{GLS}(a_s)|_{c-i}=1 
\end{equation}
It is based on  exact $x$-space derivation of  Ref. \cite{Adler:1973kz}
(compare Eq.(2.9a) with Eq.(2.9b) in  Ref. \cite{Adler:1973kz}). 

Originally  Crewther relation in the NS and vector 
channels    was  obtained 
in   Ref.  \cite{Crewther:1972kn} and analysed    in more details 
in the  work of  
Ref.  \cite{Adler:1973kz}  by means of double application  of the    
operator product expansion (OPE)  
approach for the vector-vector-axial-vector   3-point function  
\begin{equation}
\label{TV}
T_{\mu\alpha\beta}^{abc}(x,0,y)=
<0|T(V_{\mu}^a(x)V_{\alpha}^b(0)A_{\beta}^c(y))|0>~~~
\end{equation}
where $V_{\mu}^{a}=\overline{\psi}(x)\gamma_{\mu}(\lambda^a/2)\psi(x)$ and 
$A_{\beta}^c=\overline{\psi}(x)\gamma_5\gamma_{\beta}(\lambda^c/2)\psi(x)$.
In the momentum space the amplitude of Eq.(\ref{TV}) can be written 
down as 
\begin{equation}
\label{TMS}
T_{\mu\alpha\beta}^{abc}(p,q)=i\int<0|TV_{\mu}^a(x)V_{\alpha}^b(0)
A_{\alpha}^{c}(y)|0>e^{ipx+iqy}dxdy=d^{abc}T_{\mu\alpha\beta}(p,q).
\end{equation}
Thanks to the studies of   Ref.\cite{Schreier:1971um} it  
was understood  that  in the conformal-invariant limit 
this 3-index tensor is proportional to the fermion triangle one-loop 
graph, constructed from massless fermions, namely 
\begin{equation}
\label{TAF}
T_{\mu\alpha\beta}^{abc}(x,0,y)=d^{abc}N\Delta^{1-loop}(x,0,y) 
\end{equation}
Here $N$ is a number, which is proportional to anomalous constant , 
related to  $\pi^0\rightarrow \gamma\gamma$ decay \cite{Bell:1969ts}, 
 \cite{Adler:1969gk}.       
Fixing $N=1$ and keeping in mind  this  physical relation to the amplitude of 
$\pi^0\rightarrow \gamma\gamma$ decay,   we  define the coefficient function 
of the Adler D-function in the vector channel   
by taking the limit of equal charges of all quarks 
flavours and  re-writing  Eq.(\ref{DV}) as 
\begin{equation}  
D_{A}^{V}(a_s)=d_R\bigg(\sum_F Q_F^2\bigg)C_{A}^{V}(a_s)
\end{equation}
where 
\begin{equation}
C_A^{V}(a_s)=C_A^{NS}(a_s)+\frac{\bigg(\sum_F Q_F \bigg)^2}
{d_R\bigg(\sum_F Q_F^2\bigg)} \sum_{n\geq 3} d_n^{SI}a_s^n 
=C_A^{NS}(a_s)+\frac{N_F}{d_R}\sum_{n\geq 3}d_n^{SI}a_s^n~~.
\end{equation} 
The coefficient $d_3^{SI}$ is known 
from the analytical calculations of Refs.
\cite{Gorishnii:1990vf},\cite{Surguladze:1990tg},\cite{Chetyrkin:1996ez}  
 and reads 
\begin{equation}
d_3^{SI}=d^{abc}d^{abc}d_{3,1}^{SI}=
d^{abc}d^{abc}\bigg(\frac{11}{192}-\frac{1}{8}\zeta_3\bigg)~~~.
\end{equation}

Consider now the generalizations of Crewther 
relation in the channels of vector  and NS   quark currents.
They can be defined   as  
\begin{equation}
\label{CVSB}
C_A^{V}(a_s(Q^2))\times C_{GLS}(Q^2)=1+\Delta_{csb}^{V,GLS}(a_s(Q^2))
\end{equation}
and 
\begin{equation}
\label{CSBjp}
C_A^{NS}(a_s(Q^2))\times C_{Bjp}^{NS}(a_s(Q^2))=1+\Delta^{NS}_{csb}(a_s(Q^2))~~~.
\end{equation}
The expression for the conformal symmetry breaking 
contribution into Eq.(\ref{CSBjp}) can be presented 
in the following form 
\begin{equation}
\label{FactNS}
\Delta^{NS}_{csb}(a_s)=\frac{\beta(a_s)}{a_s}\bigg[K_1^{\overline{MS}}a_s+
K_2^{\overline{MS}}a_s^2+K_3^{\overline{MS}}(a_s^3)+O(a_s^4)\bigg]~~~.
\end{equation} 
where  the
coefficients $K_1^{\overline{MS}}$ and $K_2^{\overline{MS}}$ 
were first determined  in Ref. \cite{Broadhurst:1993ru}, while 
the analytical expression for $K_3^{\overline{MS}}$ was obtained 
only recently \cite{Baikov:2010je}. Note, that the possibility  that  
the factorisation of the renormalization group  
$\beta$-function in  Eq.(\ref{FactNS}) is valid 
in all orders of perturbation theory was first studied  in Ref.      
 \cite{Gabadadze:1995ei} in the momentum space. These studies  
were made by double  application of  OPE  approach to 
the axial vector-vector-vector triangle amplitude
of Eq.(\ref{TMS}) 
in the kinematic limit 
$|p^2|>>|q^2|\rightarrow \infty$ \cite{Gabadadze:1995ei}
The validity of this factorizable feature
of the generalization of  Crewther relation  
was proved in all orders of perturbation theory 
 in the coordinate 
space   \cite{Crewther:1997ux} by means of double  application of  
the   OPE to this axial-vector-vector-vector amplitude of 
Eq.(\ref{TV}) in the 
limit $x<< y$, $x,y\rightarrow 0$.

Let us  return to the problem of determination of the  
$a_s^4$-coefficient  in  Eq.(\ref{CASI}).
Its  obvious diagrammatic representation allowed  to express it in the 
following form \cite{Baikov:2010iw}: 
\begin{equation} 
\label{CBI}
d_4^{SI}=d^{abc}d^{abc}\bigg(C_F~ d_{4,1}^{SI}+C_A~ d_{4,2}^{SI}+
T_FN_F~d_{4,3}^{SI}\bigg)
\end{equation} 
It is known, that when the 
proportional to $C_F^{k}a_s^{k}$ ( with ($k \geq 1$))
contributions to $C_A^V$ and $C_{GLS}$    
are only  considered , the corresponding expression  for     
Eq.(\ref{CDI}) is valid in all orders of perturbation theory
\cite{Adler:1973kz},\cite{Kataev:2008sk}. 
Indeed, when  charge renormalization  
effects, and thus the conformal symmetry breaking contribution 
$\Delta_{csb}(a_s(Q^2))$ in the r.h.s. of Eq.(\ref{CVSB}) and 
Eq.(\ref{CSBjp})  are not taken into account, 
the conformal symmetry is effectively  restored and 
the cancellation of these  special contributions 
to   Adler functions and DIS sum rules  holds in all orders of perturbation 
theory \cite{Adler:1973kz},\cite{Kataev:2008sk}. 
In the case of Eq.(\ref{CDI}),  the cancellations 
of the similar contributions to the SI parts of
$C_A^{V}$ and $C_{GLS}$ were already observed at the $a_s^3$-level 
\cite{Broadhurst:1993ru}. In Ref.\cite{Baikov:2010iw} the  
conformal-invariant limit and   
Eq.(\ref{CDI}) were non-obviously  used  to get  the expression for 
$d_{4,1}^{SI}$ 
term from the SI order $a_s^3$ and $a_s^4$ contributions 
to Eq.(\ref{GLSSI}). Indeed, re-writing    
the variant of  expression of Eq.(\ref{CDI}) as  
\begin{equation}
\label{AV}
C_A^{V}(a_s)|_{c-i}=1/C_{GLS}(a_s)|_{c-i}~~~.
\end{equation}
one can get 
the prediction for $d_{4,1}^{SI}$ from  
Ref.\cite{Baikov:2010iw}, namely   
\begin{equation} 
\label{d41}
d_{4,1}^{SI}=-\frac{3}{2}c_{3,1}^{SI}-c_{4,1}^{SI}=-\frac{13}{64}
-\frac{1}{4}\zeta_3+\frac{5}{8}\zeta_5
\end{equation}
where the factor  $-\frac{3}{2}$ before $c_{3,1}^{SI}$  comes from the 
well-known term  $c_1=-(3/4)C_F$, which enters into 
the  contribution    
$2c_1c_{3,1}^{SI}$ to the  $O(a_s^4)$-coefficient  of    
$1/C_{GLS}(a_s)$-function in the r.h.s. of Eq.(\ref{AV}).

Let us now consider the status of  the  statement of  Ref.\cite{Kataev:1996ce}   
that for the  generalizations of Crewther relation of Eq.(\ref{CSBjp}) and 
Eq.(\ref{CVSB})  following identity holds  
\begin{equation}
\label{VGLS}
C_A^{V}(a_s(Q^2))\times C_{GLS}(a_s(Q^2))=
C_A^{NS}(a_s(Q^2))\times C_{Bjp}^{NS}(a_s(Q^2))=
1+\Delta^{NS}_{csb}(a_s(Q^2))~.
\end{equation}
The second product in Eq.(\ref{VGLS})  
was identified in Ref.\cite{Kataev:1996ce} with the product of the functions, 
which appeared  after  
application to this  triangle amplitude of OPE  in the 
limit $|q^2|>>|p^2|\rightarrow \infty$. Note, that 
during more rigorous coordinate-space studies, performed 
in Ref. \cite{Crewther:1997ux},  it was mentioned, that the  
expression for $\Delta_{csb}^{V,GLS}$ in Eq.(\ref{CVSB}) should have the 
same all-order structure as    Eq.(\ref{FactNS}), but with unfixed 
from theory coefficients $K_i^{\overline{MS}}$.  
This statement was obtained  after    
application of the OPE  
formalism to  the  triangle  Green function  of Eq.(\ref{TV}), 
in the kinematic regime  $y<< x$, $y,x\rightarrow 0$.  

In spite of the  careful  assumption  of  
Ref. \cite{Baikov:2010iw}, that the analogy of the expression 
of Eq.(\ref{FactNS}) in Eq.(\ref{CVSB}) may contain additional 
singlet-type contribution to $K_3^{\overline{MS}}$, namely $K_3=
K_3^{\overline{MS}}+K_3^{SI}$, we expected   that the  
changes of  limits in applications of 
 OPE to the same   
Green function will not lead to modification of   the concrete expression of  
the $\overline{MS}$-scheme  
coefficients from   Eq.(\ref{FactNS}) for the generalized Crewther relation 
of   Eq.(\ref{CVSB}).

Using this   expectation  and the  results for   $c_{4,2}^{SI} $ 
and $c_{4,3}^{SI}$  
obtained  in  Ref.\cite{Baikov:2010iw},    
we get   the prediction  for  two remaining  
still unknown contributions to  $d_4^{SI}$ in Eq.(\ref{CBI}).  
Taking into account  the obtained in Ref.\cite{Baikov:2010iw} prediction 
of Eq.(\ref{d41}), which  
is based on the concept of the conformal symmetry, we get        
the following  prediction  for the 
$a_s^4$ coefficient  in the expression for 
the  SI part of Adler function, namely:       
\begin{eqnarray}
\label{4SI}
&&d_4^{SI}=d^{abc}d^{abc}\bigg[ 
C_F \bigg(-\frac{13}{64}-\frac{\zeta_3}{4}+\frac{5\zeta_5}{8}\bigg) 
\\ \nonumber
&&+C_A\bigg(\frac{481}{1152}-\frac{971}{1152}\zeta_3+\frac{295}{576}\zeta_5-\frac{11}{32}
\zeta_3^2\bigg)+(T_FN_F)\bigg(-\frac{119}{1152}+\frac{67}{288}\zeta_3
-\frac{35}{144}\zeta_5+\frac{1}{8}\zeta_3^2\bigg)\bigg]
\end{eqnarray}
It is necessary to stress once more, that our prediction 
is based on the expectations that 
Crewther relations in the NS  and vector quark 
channels may have the same structure.  Both  follow  from the same axial vector-vector-vector   
triangle Green function  after change of limits of  applications of OPE 
approach  to the same amplitude, which depends from    
two conjugated  sets of variables (either  $(x,y)$ or  $(p,q)$).

Therefore, 
the direct analytical evaluation of $d_4^{SI}$-coefficient 
is  very important for getting better  understanding of the theoretical  status of
Crewther relation in different channels and   
for the  clarification of the  fundamental theoretical 
features of applications of both OPE and renormalization-group formalism 
for the  triangle amplitudes with two variables.

{\bf Note added}

After the results were presented and discussed at ACAT2011 
Workshop (Brunel Univ., London, UK, 5-9 September) and this  work 
was submitted for publication,  I became aware 
that the direct diagram-by-diagram calculations 
of $d_4^{SI}$-term were completed and presented at RADCOR2011 Workshop  
\cite{ChBKR}. The results of these calculations 
confirmed the predicted in Eq.(\ref{4SI}) coefficients of the $\zeta_3^2$-
terms, which enter into $d_{4,2}^{SI}$ and $d_{4,3}^{SI}$-terms of 
Eq.(\ref{CBI}). It should be stressed, that the analytical expressions 
for these contributions result from   from the similar expressions  for   
$c_{4,2}^{SI}$ 
and $c_{4,3}^{SI}$, obtained in Ref.\cite{Baikov:2010iw}, 
and from the original Crewther relation, obtained from Eq.(\ref{CVSB}) in the 
conformal invariant limit  after nullification of  
$\Delta_{csb}^{V,GLS}(a_s(Q^2))$-term. This is 
the second example, when the proportional to $\zeta_3$ transcendental 
term enters into the respecting conformal symmetry parts of    
order $a_s^4$ contributions into   the Adler 
$D_A^{V}$-function ( the  first similar  $\zeta_3$ contribution was 
discovered by direct analytical calculations  in Ref.\cite{Baikov:2010je}).
This fact demonstrates once more, that on the contrary to previous 
belief (see e.g. Ref.\cite{Bender:1976pw})   
the proportional to $\zeta_3$-terms 
can appear in respecting conformal symmetry contributions to 
perturbative series in realistic gauge models like QED and QCD 
and is not the accident  
(for complementary  discussions see  Ref. \cite{Kataev:2010tm}).    
Note also, that  the difference between other  analytical contributions into 
the result of Ref.\cite{ChBKR} and the ones, which enter into  Eq.(\ref{4SI}),
are  nullified   after application of the proposed in 
Ref. \cite{Kataev:2010du} test, based on the application of Banks-Zaks 
anzatz  $T_FN_F=(11/4)C_A$ \cite{Banks}. It  
comes from the special  condition $\beta_0(N_F)=0$ for the 
first coefficient of the QCD $\beta$-function.      
 Thus,  the Baikov-Chetyrkin-Kuhn  assumption,   
that the generalized Crewther relation in the channel 
of vector currents may  receive additional singlet contribution 
\cite{Baikov:2010iw},   which in 
this order of perturbation theory is  proportional 
to the first coefficient of the QCD $\beta$-function and has 
the form $\beta_0 K_3^{SI}a_s^4$, is correct.

{\bf Acknowledgements.}
I am  grateful to P.A. Baikov and K.G. Chetyrkin and V.A. Rubakov for 
useful discussions. The additional comments to this work were added  
after the invited seminar in Dep. of Physics of Univ. Santiago 
de Compostela, Spain. I wish to thank J. Sanchez Guillen for 
discussions and G.Parente for invitation.  
This work is supported in part by the 
RFBR grants  No 11-01-00182 and No 11-02-00112 and the grant NS-5525.2010.2.

\end{document}